\documentclass{aastex631}

\usepackage[T1]{fontenc}

\DeclareRobustCommand{\VAN}[3]{#2}
\let\VANthebibliography\thebibliography
\def\thebibliography{\DeclareRobustCommand{\VAN}[3]{##3}\VANthebibliography}

\usepackage{graphicx}	
\usepackage{amsmath}	
\usepackage{amssymb}	
\usepackage{amsfonts}
\usepackage[arrowdel]{physics}
\usepackage{graphicx}
\usepackage{float}
\usepackage{color, graphics}
\usepackage{verbatim}   
\usepackage{subfigure}  
\usepackage{acronym}
\usepackage{mathrsfs}
\usepackage{multirow}
\usepackage{cancel}
\usepackage{url}
\usepackage{slashed,afterpage}
\usepackage{units}
\usepackage{enumerate}  
\usepackage{mathtools}
\hypersetup{
    colorlinks=true,
    linkcolor=blue,
    filecolor=magenta,      
    urlcolor=cyan,
}
\def\beq{\begin{equation}\begin{aligned}}
\def\eeq{\end{aligned}\end{equation}}
\usepackage{tcolorbox}
\usepackage{graphicx}
\usepackage[english]{babel}
\usepackage{newtxtext,newtxmath}

\begin{document}


\title[Formation of \textit{Late Forming} PBH]{Formation and Abundance of \textit{Late Forming} Primordial Black Holes as Dark Matter}

\author{Amlan Chakraborty}
\affiliation{Indian Institute of Astrophysics, Bengaluru, Karnataka 560034, India}
\email{amlan.chakraborty@iiap.res.in}

\author{Prolay K Chanda}
\affiliation{Department of Physics, University of Illinois at Chicago, Chicago 60607, USA}
\email{pchand31@uic.edu}

\author{Kanhaiya Lal Pandey}
\affiliation{Indian Institute of Astrophysics, Bengaluru, Karnataka 560034, India}
\email{kanhaiya.pandey@iiap.res.in}

\author{Subinoy Das}
\affiliation{Indian Institute of Astrophysics, Bengaluru, Karnataka 560034, India}
\email{subinoy@iiap.res.in}

\begin{abstract}
We propose a novel mechanism where Primordial Black Hole (PBH) dark matter is formed much later in the history of the universe
between the epoch of Big Bang Nucleosynthesis (BBN) and Cosmic Microwave Background (CMB) photon decoupling. In our setup, one does not need to modify the scale-invariant inflationary power spectra; instead, a
late phase transition in strongly interacting fermion-scalar fluid (which naturally occurs around red-shift $ 10^6 \leq  \, z_{\scriptscriptstyle T} \, \leq 10^8$ ) creates an
instability in the density perturbation as sound speed turns imaginary. As a result, the dark matter perturbation grows exponentially
in sub-Compton scales. This follows the immediate formation of early dense dark matter halo, which finally evolves into PBH
due to cooling through scalar radiation. We calculate the variance of the density perturbations and PBH fractional abundances
$f(M)$ by using a non-monochromatic mass function. We find the peak of our PBH mass function lies between $10^{-16} - 10^{-14}$ solar
mass for $ z_{\scriptscriptstyle T}  \simeq 10^6$, and thus it can be the entire dark matter of the universe. In PBH formation, one would expect
a temporary phase where an attractive scalar balances the Fermi pressure. We numerically confirm that such a state indeed exists, and we find the radius and density profile of the temporary static structure of the dark matter halo, which finally evolves to PBH due to
cooling through scalar radiation.

\end{abstract}



\section{Introduction}
\label{sec:intro}

Despite decades of search, the exact nature of dark matter is still a mystery. 
The fact that no dark matter particles have been observed by direct, indirect, or collider searches casts a doubt \cite{}on the assumption that dark matter is just a weakly interactive massive particle (WIMP) motivated by electro-weak scale physics. Once we start to explore possible DM candidates beyond WIMP,  Primordial black holes (PBH) come as one of the strongest and most well-studied candidates. Generally, in most of the models of PBH, they are formed deep in a radiation-dominated era via the collapse of large density perturbations \cite{1967SvA....10..602Z, 1971MNRAS.152...75H}. Though PBH is a simple and well-motivated DM  candidate, the present PBH dark matter density is subject to stringent constraints from various observational and theoretical studies \cite{}. In principle, at the time of their generation, PBHs could have masses starting from the Planck mass ($10^{-5}$g) to stupendously large masses such as $\sim 10^{17} {M}_\odot$, the horizon mass at the time of matter–radiation equality \citep{2021MNRAS.501.2029C}. On a cosmological scale, PBH dark matter would behave almost like dark particle matter; however, depending on their mass, on galactic and smaller scales, it can have characteristic observable consequences. PBHs can also have implications in early black hole seeding, and formation of the first stars and galaxies, accretion onto massive enough PBHs can account for the detected X-ray and infrared backgrounds and their cross-correlation \citep[and references therein]{2022ApJ...926..205C}. Although low mass PBHs are expected to evaporate via Hawking radiation faster, those with initial mass $\gtrsim 10^{15}$ g have a lifetime longer than the age of the Universe \citep{1974Natur.248...30H,1975CMaPh..43..199H}.

There have been extensive efforts to search for the PBHs using various observations. These include observations of the extragalactic $\gamma$-ray background, gravitational microlensing experiments (e.g., OGLE-I-IV, using Kepler objects, Eridanus-II star clusters), cosmic microwave background experiments, dynamical constraints, and accretion constraints \citep[and the references therein]{PhysRevLett.123.251101,PhysRevD.101.123514,PhysRevLett.125.101101,2020ARNPS..70..355C,2021RPPh...84k6902C} etc. More recently, gravitational wave astronomy has opened up a new avenue for the search of primordial black holes through the gravitational waves generated either by their coalescence or associated with their generation \citep[and references therein]{2016PhRvL.116t1301B,2016PhRvL.117f1101S,2018CQGra..35f3001S,2020PhRvD.101l3535K,2021ApJ...910L...4K,2021JCAP...03..068H}. 
Fig.~10 in \cite{2021RPPh...84k6902C} gives a nice overview of the current status of the PBH abundance constraints over all the possible mass ranges. From the figure one can find that the potential allowed mass range over which PBHs can still make up for the entire dark matter belongs to  the $10^{17}$g - $10^{24}$g ($10^{-16} {M_\odot}$ - $10^{-10} {M_\odot}$).   

Primordial black holes are generally produced from the collapse of inhomogeneities in the early universe, unlike the stellar black holes which form from the collapse of a star. To create such extreme inhomogeneity in a very early epoch, one generally depends on modification in inflationary power spectra \cite{carr_inf}, \cite{iv_inf}, \cite{2018CQGra..35f3001S}, \cite{Ashoorioon2018}, \cite{Bhattacharya_2021} in the early universe, where the model permits a peak at very small length scale in the primordial power spectra \citep{1996PhRvD..54.6040G,2015PhRvD..92b3524C,2017PDU....18...47G,2017JCAP...09..020K,2018PhLB..776..345E,2018PhRvD..97b3501B,2018JCAP...03..016F, Pi_2018, Ashoorioon2021}. This implies the density contrast to be very high at those scales, and that gives the perfect environment for the formation of PBH when those modes enter horizon. 

But an alternative route is to keep the success of simple inflationary model with  scale invariant power spectra as it is  but creating high  sub-horizon scale density contrast with some phase transition that naturally take place as the universe cools down. For example, QCD PBH are formed due to the temporary  drop in the pressure  around epoch of QCD phase transition \cite{1997PhRvD..55.5871J,1998astro.ph..8142W,1998PhR...307..155J,2005PhRvD..71h7302H,chris_byrnes}.However since the QCD PBHs will have masses of the order $\sim 0.1-10$ $M_\odot$, they can not account for all the dark matter given the current constraints on $f(M)$ for these masses (see Fiq.\ref{fig:fm_plt_exp}). Other first order phase transition  driven by  bubble nucleation or bubble collision can also create PBHs \cite{45_nature}, \cite{PhysRevD.26.2681},\cite{25f71}, \cite{PhysRevD.50.676}, \cite{1998AstL...24..413K}, \cite{etde_20498414}, \cite{10.1143/PTP.68.1979}, \cite{PhysRevLett.125.181304}, \cite{Gross2021-nw}, \cite{baker2021primordial},\cite{KAWANA2022136791}. Existence of attractive long range fifth force \cite{Das:2021wad} or scalar field fragmentation \cite{PhysRevLett.119.031103}, \cite{PhysRevD.96.103002}, \cite{PhysRevD.98.083513}, \cite{Cotner_2019} can also lead to PBH  dark matter formation in pre-BBN era. Once PBHs form through above process, typically the  PBH mass is of the order of  horizon size  at the  formation epoch. Thus knowing the formation epoch  more or less fixes the PBH mass.  If the PBHs form during radiation dominated era, the PBH mass, $M_{\rm BH}$, differs from the horizon mass, $M_{\rm H}$, by only a factor of unity $\gamma\lesssim 1$ \citep{2021RPPh...84k6902C},
\beq\label{eq:MBH2Tnt}
M_{\rm BH} = \gamma M_{\rm H} =  \frac{\gamma c^{3}t}{G}\approx 2.03\times 10^{5}\gamma \left(\frac{t}{1~{\rm s}}\right)M_{\odot}.
\eeq

But in this  work, we break this simple relation between PBH mass and formation red-shift by introducing a new scale in 
the formation process.
The main goal here is to form PBH at a later epoch (between BBN and  CMB decoupling epoch ) but with the PBH masses much lesser than horizon mass at the time of formation. In standard $\Lambda$CDM cosmology, it is not possible to form PBH  during this epoch as gravity is not strong enough to overcome radiation pressure. 
But the presence of a new additional force stronger than gravity can make the story different \citep{PhysRevD.100.083518,Das:2021wad}. For example, recently, it was pointed out \citep{PhysRevD.100.083518} that an early dense DM halo can form deep in a radiation-dominated era when an attractive scalar force makes the dark matter perturbation grow much faster $\delta \sim a^p$ than standard logarithm growth.  As a follow-up work on this \citep{kushenko}, it was further demonstrated that these early halos could lose energy through scalar radiation and can indeed form PBH  as a viable dark matter candidate. 

Here we propose a scenario where the growth rate of primordial inhomogeneities of a fermionic particle coupled with a scalar field can even grow much faster than polynomial. We show that exponential growth is well possible in a scalar-fermion interaction \cite{Gogoi:2020qif}. This exponential growth occurs when our $\rm{keV}$mass fermionic particle turns non-relativistic around $z_{\scriptscriptstyle{T}} \sim 10^6$ and sound speed of DM perturbation turns imaginary \cite{Afshordi:2005ym}. 
We derive the matter power spectrum for the above mentioned exponential growth scenario and do a detailed calculation of the PBH mass function in this scenario using the scaling relation for a critical gravitational collapse of a massless scalar field given by \cite{1993PhRvL..70....9C} and the Press-Schechter formalism \cite{PSFormalism}. We find that for the $ z_{\scriptscriptstyle{T}} \simeq 10^6$ the peak of the PBH mass function lies between $10^{-15} - 10^{-12} M_\odot$ and therefore it can account for almost entire dark matter in the universe ($f(M)\simeq1$) given the current constraints on the PBH abundances.

The plan of the paper is as follows. 
We describe the growth of density perturbation of the fermion-scalar fluid  in section \ref{sec:lfpbh}. In the next section \ref{sec:scl_rad}, we numerically solve for the static structure of the primordial halos. Section \ref{sec:mass_fn_calc} provides the adopted formalism for the derivation of mass function for Late forming PBH  followed by the main numerical results. In section \ref{sec:conclusion}, we summarize the main findings and future perspective.

\begin{figure}
\centering
     \includegraphics[width=0.6\columnwidth]{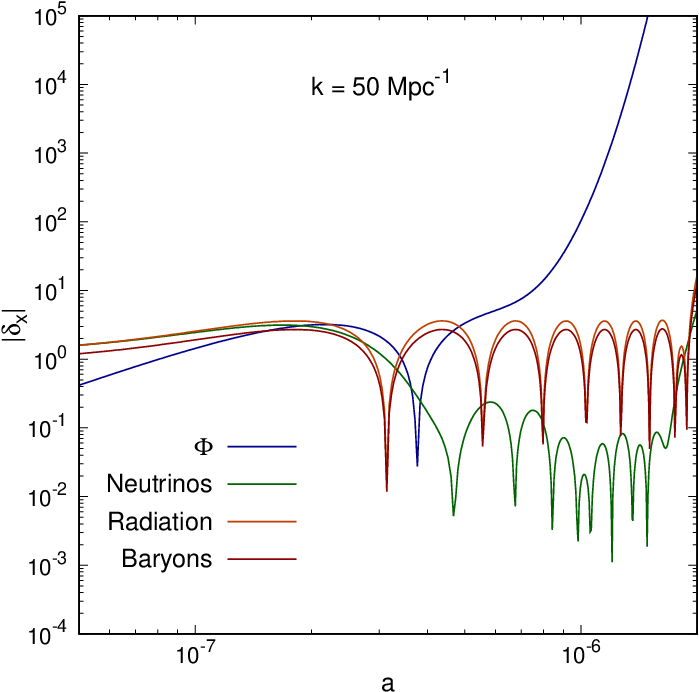}
     \caption{$|\delta|$ vs scale factor $a$ for a typical wave number which is inside the Compton scale of the scalar field around the PBH formation redshift $z_{\scriptscriptstyle T} = 10^6$. We can see that the growth of the $|\delta_\Phi|$ is exponential which makes this mode go non-linear and finally collapse to form PBHs.}
     \label{fig:absdel_vs_a}
\end{figure}

\section{Late Forming PBH: The mechanism}
\label{sec:lfpbh}

This paper proposes a novel mechanism for forming PBH at a late epoch ( around $z \simeq 10^4 - 10^7$). But later, we will see if we want PBH to make up for the entire DM from cosmological constraints $ z \geq 10^6$. The key ingredient is --
a dark matter fermion that couples to a long-range scalar. Due to the presence of coupling to matter, scalar field evolution is controlled by an effective potential $V^{\rm eff}(\phi)$ \cite{Das:2005yj}. Typically, the shape of the effective potential has a minimum, and the scalar field adiabatically follows that minima \cite{Khoury:2003aq} of the effective potential. The dynamics of scalar with a self-interacting potential in such chameleon theories can also give rise to a temporary epoch of early dark energy before CMB \cite{Karwal:2021vpk,Gogoi:2020qif} and can relax Hubble tension. It was explicitly shown for the case of usual dark energy \cite{Afshordi:2005ym} that such a system of interacting dark matter-scalar fluid would encounter a perturbative instability when matter particle turns non-relativistic. In \cite{Gogoi:2020qif}, the same mechanism was extended for early dark energy theories, and it was shown that even in radiation dominated era, the EDE phase would encounter exponential instability in dark matter perturbation when it turns non-relativistic. In our case, the DM mass is of the order of \rm{KeV}--so one would expect the instability to occur around $z_T \simeq 10^6 - 10^7$. 

To find the exact epoch when fluctuation grows non-linear,  one needs to solve for the linear perturbation equations-  evolving it from big bang to a redshift around $z_T$. We adopt the generalized dark matter(GDM) formalism for our DM-scalar fluid as done in \cite{Hu:1998kj}. In this formalism, the background equation of the fluid is parameterized by its equation of state, $w_{\phi-\psi}$ which we take to be a function of the redshift $z$. In early times, since the fluid was relativistic, the coupling between dark matter and the scalar field could be ignored ( as it couples through the trace of energy-momentum tensor). So, we take $w_{\phi-\psi_1}\sim\frac{1}{3}$ at high redshifts, and as the dark matter tends to become non-relativistic around a  redshift $z_T$, the effective coupling turns on. For a quadratic self-interacting scalar potential, the fermion-scalar fluid effectively behaves like early dark energy for a short duration - soon after, the fluid 
sound speed square $c_s^2$ turns negative, which results in a strong instability in the fluid perturbation. For details of the evolution of $c_s^2$ and $w_{\phi-\psi_1}$ we refer to the work \cite{Gogoi:2020qif,Afshordi:2005ym}. It was shown there that the DM-scalar fluid perturbation starts to grow exponentially even in radiation dominated era once the sound speed becomes imaginary. In this work, we show that this mechanism would form a dense early DM halo which finally collapses into PBH due to cooling through scalar radiation \cite{kushenko}. 

The perturbations equations for our fluid in synchronous gauge from the GDM formalism are given by \cite{Gogoi:2020qif}:
\begin{eqnarray}
\dot{\delta} &=& -(1+w)\left(\theta+\frac{\dot{h}}{2}\right)-3(c_s^2-w)H\delta \label{eq:delta_dot}\\
\dot{\theta} &=& -(1-3c_s^2)H\theta + \frac{c_s^2k^2}{1+w}\delta
\end{eqnarray}

To solve the background and the perturbation equations, we modify the Boltzmann code CLASS \citep{2011arXiv1104.2932L,2011JCAP...07..034B} to replace the CDM component with an extra fluid component which describes our fluid through $w_{\phi-\psi}$ and $c_s^2$. In 
Fig.\ref{fig:absdel_vs_a}, 
we indeed see that around $z_F$, the density fluctuations of the $\phi$-$\psi$ fluid shoots up to a huge value compared to those of the other components such CDM and neutrinos from standard $\Lambda$CDM  scenario. 
 This exponential growth makes the perturbation turn non-linear very fast, followed by the formation of dense early halos, which evolve into PBH \citep{kushenko} by loose energy through scalar radiation.
 
 There are two main concerns related to the formation of dark matter PBH at such a late epoch few \textit{e-foldings} prior to CMB. First of all, one needs to check if the appearance of dark matter so late in the universe is at all viable from CMB and structure formation perspective. From the recent studies it seems,  if DM is produced earlier than  $z \geq 10^6$, one can satisfy both  CMB \cite{Sarkar:2017vls, Agarwal:2014qca}, Large Scale Structure \cite{Sarkar:2017vls,Sarkar:2015dib,Sarkar:2017vls} observations as well as constraints from  local Milky Way satellite observations \cite{Das:2020nwc}.
 
  The second challenge is the horizon size around this late redshift is very large, and if one produces horizon size PBH, the PBH mass would be stupendously big (M $> 10^{11} {M}_\odot$, known as SLAB- the Stupendously Large Black Holes). This will be subject to many constraints arising from dynamical friction and destruction of galaxies in the cluster ${\rm f}_{\rm PBH} \lesssim 10^{-3}$ \citep{2021MNRAS.501.2029C}.
  But we will show how the mass of the scalar field in our example is typically much higher than the local Hubble constant. So the Compton wavelength or the range of the fifth force is much smaller than the horizon. As an attractive fifth force is the main reason for forming PBH- one would naturally expect the size of the PBH to be much smaller than the horizon mass at the formation epoch.

During the radiation domination, $H=(2t)^{-1}$, such that the age of the Universe $t$ can be related to the temperature of the thermal bath $T$ as the following,
\beq\label{eq:t2T}
t\approx 2.42~g_{\ast}^{-1/2}\left(\frac{T}{1~{\rm MeV}}\right)^{-2}{\rm s}.
\eeq

\begin{figure}
    \centering
    \includegraphics[scale = 0.4]{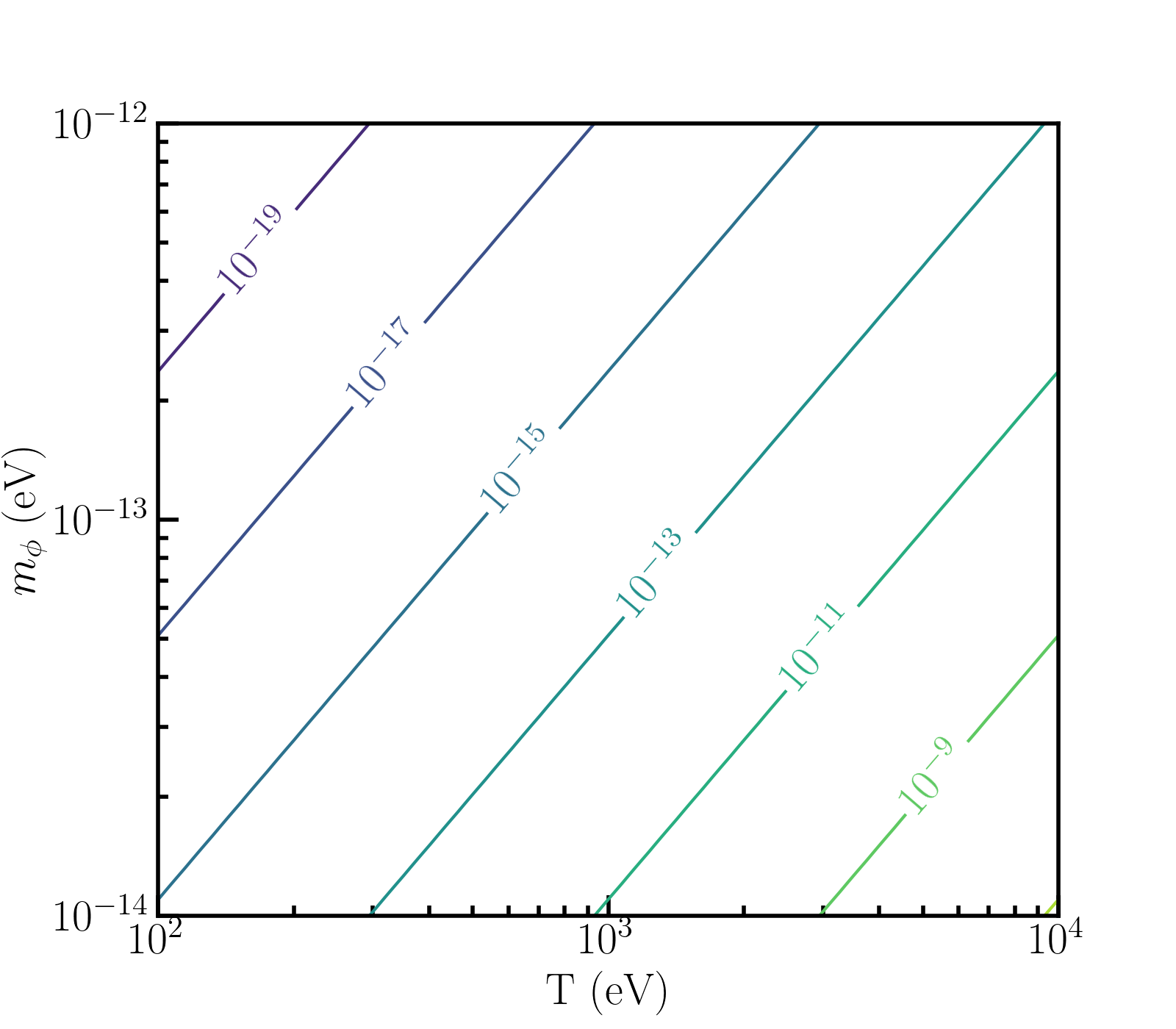}
    \caption{The contours above delineate different PBH masses in the solar units as a function of the $T-m_{\phi}$ parameter space, derived using eq.~\eqref{eq:MBH-T-mPhi} where $T$ is the temperature of the Universe at epoch of formation of PBH and $m_{\phi}$ is the mass of the scalar field mediating the fifth force. }
    \label{fig:mPhi-T}
\end{figure}

 Using eqns. \eqref{eq:MBH2Tnt}, and \eqref{eq:t2T}, we find that for such a PBH being formed at a redshift $z_T \simeq 10^6 $, is expected to have a horizon size  mass, $M_{\rm BH} \sim 10^{12}M_{\odot}$, as evaluated from eq.~\eqref{eq:MBH2Tnt} which is much heavier than the mass window mentioned above.
  But as in our case,  only a fraction of horizon mass (enclosed in the Compton volume of the scalar)  collapses into the black hole, we find 
\beq\label{eq:MBH-T-mPhi}
M_{\rm BH} = \frac{\gamma c^{3}t}{G}\approx 2.03\times 10^{5}\gamma \left(\frac{t}{1~{\rm s}}\right)M_{\odot}{\left(\frac{m_{\phi}}{H}\right)^{-3}},
\eeq
where, $m_{\phi}$ is the mass of the scalar field. The term $\left(m_{\phi}/H\right)^{-3}$ provides with an extra parameter for tweaking $M_{\rm BH}$, as can be seen from Fig.\ref{fig:mPhi-T}.

We discuss a scenario where the PBHs form at a temperature, $\sim 1$ keV in the presence of a scalar field of mass, $m_{\phi}\sim 10^{-13}$ eV, so that the factor $\left(H/m_{\phi}\right)^{3}$ corresponds to a value $\sim 10^{-25}$. The density of the thermal bath derived as the function of the Hubble parameter,  $ 3H^{2}M_{\rm Pl}^{2}/8\pi$, is evaluated as $\sim 10^{12}~{\rm eV}^{4}$ during this time period. Assuming a monochromatic mass function the masses of the PBHs formed can be estimated as the mass enclosed within the Hubble horizon, $M_{H}\sim (H^{-1})^{3}\rho\sim M_{\rm Pl}^{2}/H$, multiplied by the factor, $\left(H/m_{\phi}\right)^{3}$. With the choice of the parameter space as discussed above, the PBHs are expected to form with a mass, $M_{\rm BH}\sim 10^{-14}~M_{\odot}$. This mass corresponds to a Schwarzschild radius, $\sim M_{\rm BH}/M_{\rm Pl}^{2}\sim 10^{-9}~{\rm cm}$, far below the distance measures of this study.

\begin{figure}
    \centering
    \includegraphics[scale = 0.35]{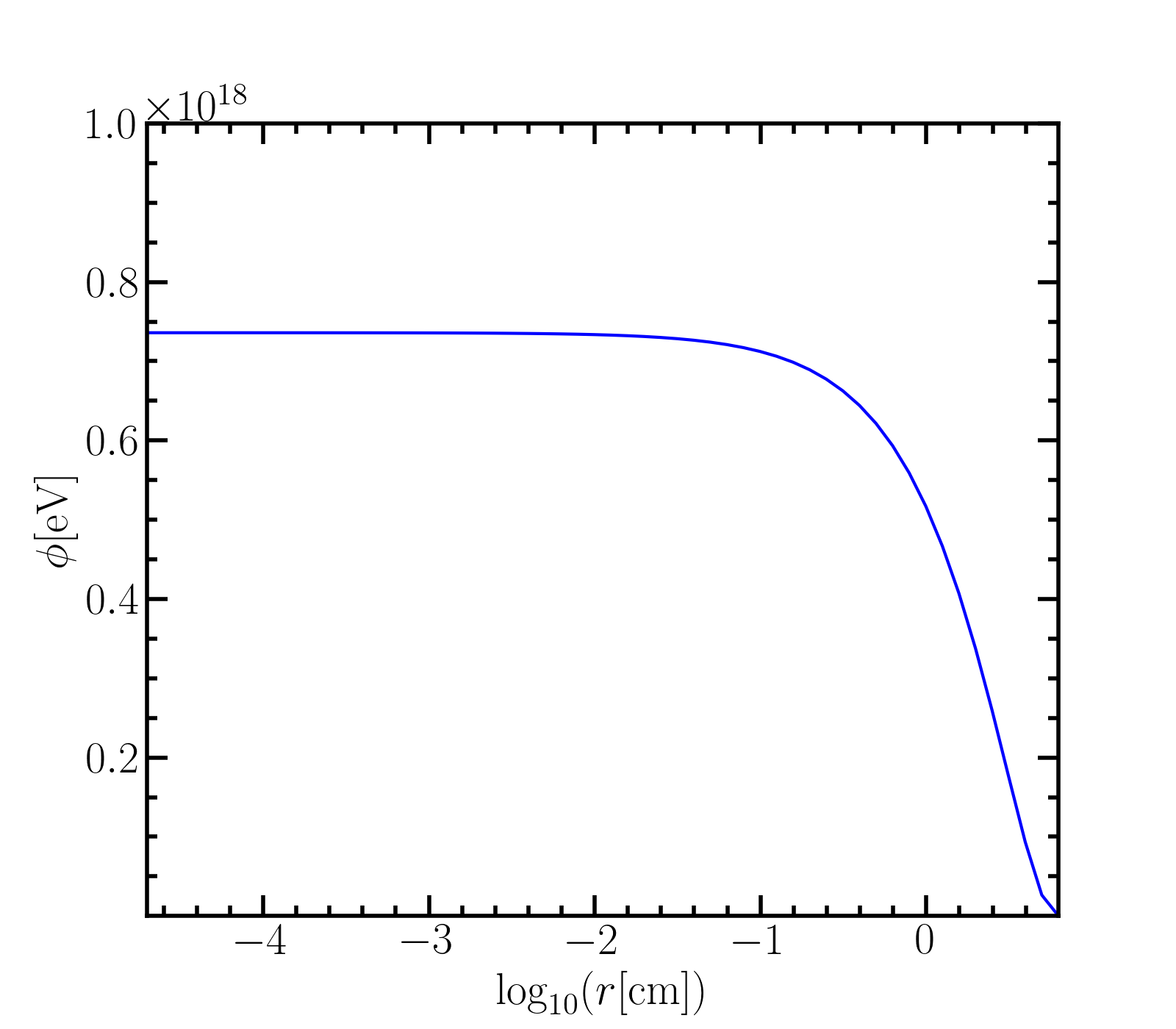}
    \includegraphics[scale = 0.35]{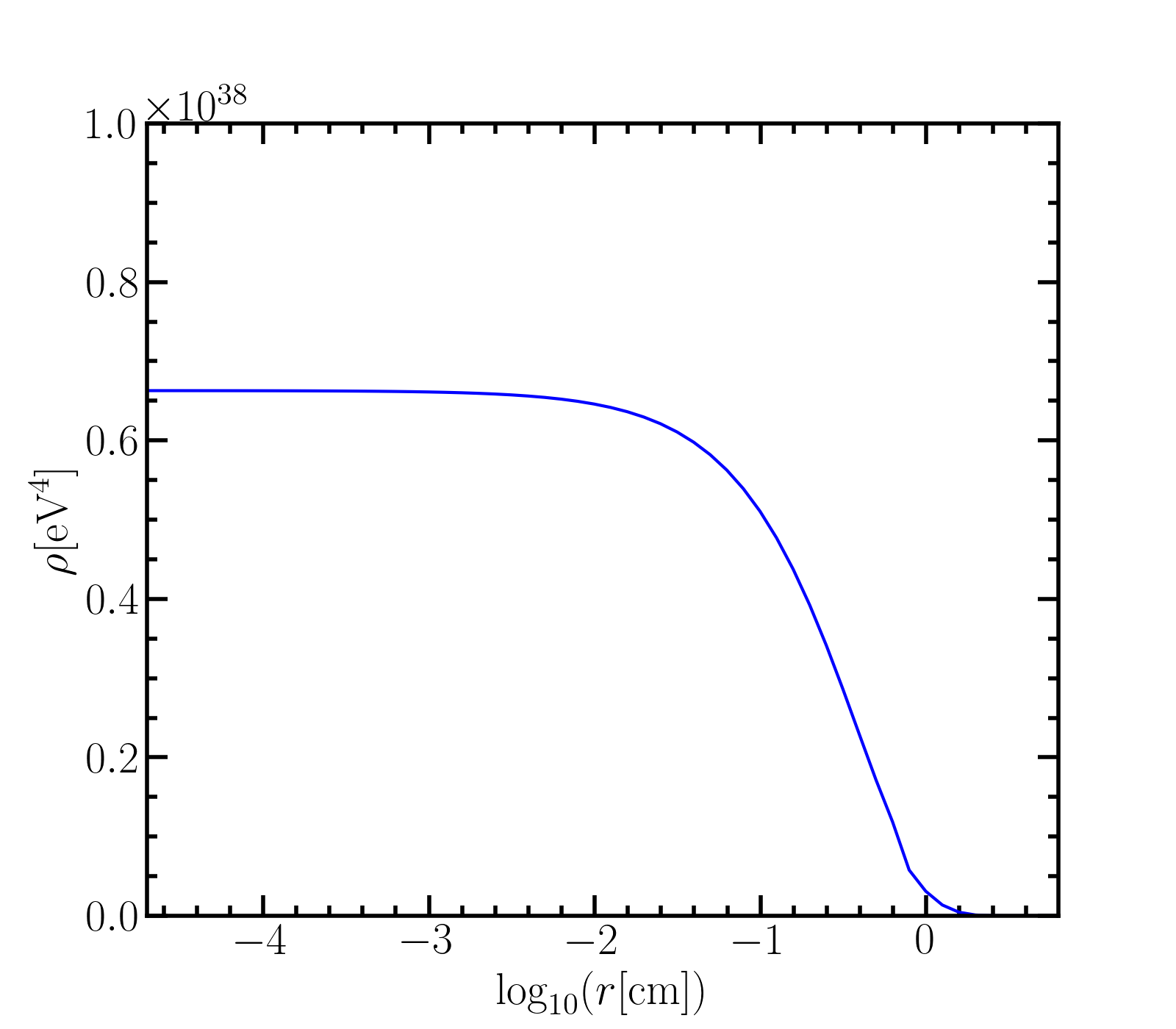}
    \caption{In the above, we demonstrate the static profile, which is obtained by solving eqns.~\eqref{eq:diffP}, and, \eqref{eq:StaticPhi}. The plot at the left shows the static profile for the scalar field, $\phi$, and the plot at the right delineates the density profile, $\rho$, of the resulted dark matter halo. As discussed in the main text, this halo may later collapse due to cooling through scalar radiation and eventually form PBH \citep{kushenko}.}
    \label{fig:phi-rho}
\end{figure}

\section{Static Structure of primordial Halo}
\label{sec:scl_rad}

  It was pointed out in \cite{kushenko} that in the presence of an attractive long-range force mediated by a scalar field, an intermediate state of a dark matter halo forms, as this attractive `fifth force' is balanced by the Fermi pressure. Accelerated dark matter particles moving in the halo will emit scalar radiation due to the long-range interaction, resulting in the halo losing energy so that it may collapse rapidly to form PBH. The dipole radiation due to the coherent motion of the particles should vanish because the particles are identical, while the higher moments are expected to be negligible. On the other hand, where the radiation is due to particles behaving as incoherent sources, the radiated power depends linearly on the number of dark matter particles in the halo. The halo can also lose energy through scalar bremsstrahlung radiation, where the dominating radiation is quadruple as the dark matter particles are identical \cite{Maxon1967,Maxon1972}.
As the halo decreases in size, the collapse timescale increases, the diffusion timescale decreases, and at some point, the diffusion becomes slower than the collapse. This results in the radiation being trapped, and the cooling happens from the surface. A dark matter halo can collapse into a black hole only if the rate of energy loss, $\tau_{\rm loss}$, is less than the expansion rate of the Universe, $\sim H^{-1}$.

In this section,  we numerically show that, indeed, the temporary static phase is possible where fermi degeneracy pressure balances scalar forces.
Numerically we solve for the static profile for the $\phi$ field. We refer to \cite{Brouzakis:2005cj,Lee:1986tr,Chanda:2017coy} for detailed derivation of these equations. We consider the dark matter to be a fermion field, $\psi$, self-interacting through a scalar field $\phi$. We discuss a more general interaction as compared to the original mass varying neutrino model \citep{Fardon:2003eh} which considers Majorana mass term to be a linear function of $\phi$. In this analysis we introduce a function $f(\phi)$ so that the dark matter mass is obtained as, $m_{\psi} = m_{D}^{2}/f(\phi)$, where, $m_{D}$ is the Dirac mass. We choose the self-interaction to have the form, $f(\phi) = \lambda_{\psi}\phi$. 

Prior to the phase transition, an effective potential controls the dynamics of the $\phi$ (\cite{Fardon:2003eh})
\begin{equation}
V_{\rm eff} = \rho_{\psi }+V(\phi).
\end{equation}

In our analysis, we use, $V(\phi) = m_{\phi}^{2}\phi^{2}$, where, $m_{\phi}$ is the scalar field mass.

 We assume the dark matter particles to be weakly interacting and non-scattering such that the motion of the $\psi$ particles can be determined using the Thomas-Fermi approximation \cite{Brouzakis:2005cj}, that is, the physical parameters such as density, pressure, and number density are characterized by the distribution,  $\sim\left[1+\exp\left(\epsilon_{F}/T\right)\right]^{-1} $, where, the Fermi energy, $\epsilon_{F} = \sqrt{p_{F}^{2}+m_{\psi}^{2}}$; $p_{F}$ being the Fermi momentum. We consider a simpler scenario by doing the calculations in the zero temperature limit \cite{Brouzakis:2005cj}, and, derive an explicit form of the trace of the energy-momentum tensor, $T_{\mu}^{\mu} = \rho - 3p$: 
\begin{align}\label{eq:EnergyMomentum}
T_{\mu}^{\mu}&=\frac{m_{\psi}^2}{2 \pi^2}\left(p_F \sqrt{p_F^2 + m_{\psi}^2} - m_{\psi}^2 \mathrm{ln}\left(\frac{p_F + \sqrt{p_F^2 + m_{\psi}^2}}{m_{\psi}}\right)\right),
\end{align}
 and the pressure
\begin{equation}\label{eq:p}
p=\frac{m_{\psi}^{2}}{8\pi^{2}}\left[\frac{2p_{F}^{3}}{3m_{\psi}^{2}}\sqrt{p_F^2 + m_{\psi}^2}  - \left(p_F \epsilon_F - m_{\psi}^2 \mathrm{ln}\left(\frac{p_F + \epsilon_F}{m_{\psi}}\right)\right) \right],
\end{equation}
where, $\epsilon_F = \sqrt{p_F^2 + m_{\psi_{1}}^2} $. 

In the weak limit of general relativity, the Klein-Gordon equation can be written as
\begin{equation}\label{eq:StaticPhi}
\phi'' + \frac{2}{r} \phi' = \frac{dV(\phi)}{d\phi} - \frac{d\mathrm{ln}[m_{\psi}]}{d\phi} T_{\mu}^{\mu},
\end{equation}
and the Euler equation for pressure, $p$, 
\begin{equation}\label{eq:diffP}
\frac{dp}{d\phi} = \frac{d\left[\ln m_{\psi}\right]}{d\phi}T^{\mu}_{\mu}.
\end{equation}

We obtain a static solution for $\phi$ as demonstrated in Fig.~\ref{fig:phi-rho}, which is determined by how the attractive fifth force is balanced by the local Fermi pressure as derived from the Euler equation, \eqref{eq:diffP}, and the Klein-Gordon equation \eqref{eq:StaticPhi}. The mass enclosed in the dark matter halo corresponds to a Schwarzschild radius of $\sim 10^{-9}$ cm, which is much smaller than the size of the halo. The density at the core of the profile can be predicted without explicitly carrying out the numerical calculations. Because of the self-interactions through the $\phi$ field, the density of dark matter particles near the core of the profile should be much denser by a factor of  $\left(m_{\phi}/H\right)^{3}$. Therefore, for PBHs formed at $\sim 1 $ keV for dark matter particles self-interacting through a scalar of mass, $m_{\phi}\sim 10^{-13}$ eV, the core density should be, $\sim 10^{12}~{\rm eV}^{4} \left(m_{\phi}/H\right)^{3}\sim 10^{37}~{\rm eV}^{4}$, which, agrees very well with the results from the numerical studies. We should remind ourselves from the earlier calculations that at $\sim 1$ keV, the factor, $\left(m_{\phi}/H\right)^{3}\sim 10^{25}$, for the given scalar mass, and, the factor $10^{12}~{\rm eV}^{4}$ is the density of the thermal bath evaluated at the same temperature.

\section{PBH mass function and PBH DM abundance}

\label{sec:mass_fn_calc}

In this section, we focus on estimating the dark matter abundance of PBH from the collapse of primordial fluctuations in the radiation-dominated era. Though we concentrate mostly on the cosmology with the exponential growth of perturbation prior to the phase transition, our treatment and numerical code developed in the current section is kept generic and can be applied for calculating  PBH abundance with different perturbations growth rate.  It is instructive to note that here in this work, we consider a non-monochromatic mass function, i.e., the mass of PBH formed from collapse does not necessarily have to be equal to the horizon mass.

The mass distribution of PBH is usually stated in terms of $f(M)$, the fraction of CDM made
up of PBHs of a given mass M, and is given by \citep{chris_byrnes},

\beq\label{eq:fM_betaM}
f(M) = \frac{1}{\Omega_{\rm CDM}} \frac{d\Omega_{\rm PBH}}{d\ln M} 
\eeq
where $\Omega_{\rm CDM}$ and $\Omega_{\rm PBH}$ are the abundances of CDM and PBH respectively.

Using Press-Schechter formalism \cite{PSFormalism} and taking into account the mass fraction of the Universe that collapses to form PBHs at the time of formation, we can write expression for $\beta (M)$ as,
\beq\label{eq:betaM}
\beta(M) = 2 \int_{\delta_{\rm c}}^\infty \frac{M}{M_h} Pr(\delta) \ d\delta
\eeq
where $\delta_{\rm c}$ is the critical value of $\delta$ at the time of horizon entry for PBH formation, and $Pr(\delta)$ is the probability density function of the density contrast. Assuming a Gaussian distribution of density fluctuations, $Pr(\delta)$ can be given as,

\beq\label{eq:P_del}
Pr(\delta) = \frac{1}{\sqrt{2\pi \sigma^2(R)}}\exp({-\frac{\delta^{2}}{2\sigma^2(R)}}) 
\eeq
where $\sigma(R)$ is smoothed density fluctuations over a smoothing radius $R$ and can be given as, 

\beq\label{eq:std}
\sigma^{2}(R)=\int_{0}^{\infty}\frac{dk}{k}\Delta(k) W_{R}^{2}(k)
\eeq
here $W_{R}^{2}(k)$ is the window function of smoothing radius $R$ which we take as a step function, is given by,
\beq\label{eq:shrp_k}
W_{R}(k)=\theta\left(\frac{1}{R}-k\right)
\eeq
The smoothing radius $R$ can be related to horizon mass $M_h$ using the following relation, (see, Appendix~\ref{a_1})
\beq\label{eq:xto_m_h}
\frac{1}{R}=\frac{k_B T_0}{2}\left(\frac{g_{*,s}(T_0)}{g_{*,s}(T_h)}\right)^{1/3} \left(\frac{g_{*,\rho}(T_h)}{45\hbar^{3}c^3 \pi G}\right)^{1/4}\frac{1}{\sqrt{M_h}} 
\eeq
The dimensionless power spectra $\Delta(k)$ in Equation~\ref{eq:std} can be written as,
\beq\label{eq:dim_pow_spec}
\Delta(k,z)= \frac{k^{3}P(k,z)}{2\pi^{2}}= A_s \delta^2(k,z) 
\eeq
where, $P(k)$ is the density power spectra and $A_s$ is the primordial power spectrum amplitude.

Now considering the formation of PBH from the gravitational collapse of primordial density fluctuations in the radiation dominated phase of the early universe as a critical phenomenon, we can write the following equation for the PBH mass at its formation \citep{1993PhRvL..70....9C},
\beq\label{eq:pbh_mass}
M=KM_h(\delta-\delta_c)^\gamma
\eeq
where $M$ is the mass of the primordial black hole; $\delta$ is the density contrast, and $\delta_{\rm c}$ is the critical value of $\delta$ at the time of horizon entry, $M_{h}$ is the horizon mass at the time the fluctuation entered the horizon, $K$ and $\gamma$ are constants which depend on the shape of the fluctuations and background equation of state respectively.

The above equation can be inverted to give $\delta$ as a function of $M$ as,
\beq\label{eq:del_as_M}
\delta = \left( \frac{M}{K M_h} \right) ^{1/\gamma} + \delta_{\rm c} = \mu^{1/\gamma} + \delta_{\rm c}
\eeq
where, $\mu=M/(K M_h)$. Now using Eq~\ref{eq:fM_betaM}, \ref{eq:betaM} and \ref{eq:del_as_M} we can write $f(M)$ as \citep{chris_byrnes},

\beq\label{eq:fM}
f(M)= \frac{2}{\Omega_{\rm CDM}}\int_{-\infty}^{\infty} Pr(M_h) \frac{M}{\gamma M_h}\mu^{1/\gamma}\left(\frac{M_{\rm eq}}{M_h}\right)^{1/2}d\ln{M_{h}}
\eeq
where $Pr(M_h)$ is the probability density function of the density contrast in terms of $M_h$,
\beq\label{eq:Pr_Mh}
Pr(M_h) = \frac{1}{\sqrt{2\pi \sigma^2(M_h)}}\exp({-\frac{(\delta_{\rm c}\;+\;\mu^{1/\gamma})^{2}}{2\sigma^2(M_h)}}) 
\eeq

\subsection{Power spectrum $P(k)$ for exponentially growing $\delta$} \label{sec: exp}
In this section, we discuss the case where the density contrast grows exponentially, i.e., $\delta\propto \exp({\lambda k c_s t})$, where $c_s$ is the sound speed, and $\lambda$ is a constant and has been tuned to a suitable value such that the formed PBH fractional abundances $f(M)$ are within 1.
In this case, in which the scaling (the exponential growth) regime starts from $z_{\rm in}$ and lasts till redshift $z_{\scriptscriptstyle T}$, the power spectrum can be written as,
\beq\label{eq:pow_spec_exp}
P(k,z) = \begin{cases*} \frac{2 \pi^2 A_s}{k^3} & if $k\leq k(z_{\scriptscriptstyle T})$, \\
                             \frac{2 \pi^2 A_s}{k^3} \exp(\frac{2\; c_m}{(1+z_{\scriptscriptstyle T})^2})\exp(-\frac{2\; c_m}{(1+z(k))^2}) & if $k(z_{\scriptscriptstyle T})<k\leq k(z_{\rm in})$, \\
                             \frac{2 \pi^2 A_s}{k^3} \exp(\frac{2\; c_m}{(1+z_{\scriptscriptstyle T})^2}) \exp(-\frac{2\; c_m}{(1+z_{\rm in})^2}) \left(\frac{\log(1+z_{\rm in})}{\log(1+z(k))}\right)^2 & if $k>k(z_{\rm in})$
\end{cases*}
\eeq
where $c_m= {\lambda k c_s}/{2 \sqrt{\Omega_{r,0}}H_0}$ and 
where $z$ as a function of $k$ can be given as, (see, Appendix~\ref{a_2})
\beq\label{eq:k(z)_and_z(k)_2}
1+z(k) = \frac{k}{k_B T_0 }\left(\frac{g_{*,s}(T_0)}{g_{*,s}(T_h)}\right)^{-1/3} g_{*,\rho}^{-1/4}(T_h) \left(\frac{4\pi^3 H_0^2 \,G\, \Omega_{r,0}}{45\,\hbar^3 c^9}\right)^{-1/4}
\eeq
where, $k(z_{\scriptscriptstyle T})$ and $k(z_{\rm in})$ are different wave numbers that just entered the horizon corresponding to their respective redshifts,
$z_{\rm in}$ is the redshift at which this growth starts and ends at a redshift $ z_{\scriptscriptstyle T}$. In the equation \ref{eq:pow_spec_exp} the first case ($k\leq k(z_{\scriptscriptstyle T})$) represents the scales which are still outside the horizon at the redshift $ z_{\scriptscriptstyle T}$, the second case ($k( z_{\scriptscriptstyle T})<k\leq k(z_{\rm in})$) represents the scales that have entered during the exponential growth period, and the last case ($k>k(z_{\rm in})$) represents the scales which entered before the growth period and grew logarithmically in the initial phase following the standard $\Lambda$CDM cosmology driven growth. After that all the scales grow by equal amount in the exponentially growing regime. For our calculation we have taken $z_{\rm in}= 3 \times 10^7$.
We cut the power spectrum for $k > k(z_{\rm in})$ as these modes enter the horizon before the exponential growth regime starts.

\subsection{Results}

\begin{figure}
\centering
\includegraphics[width=0.75\columnwidth]{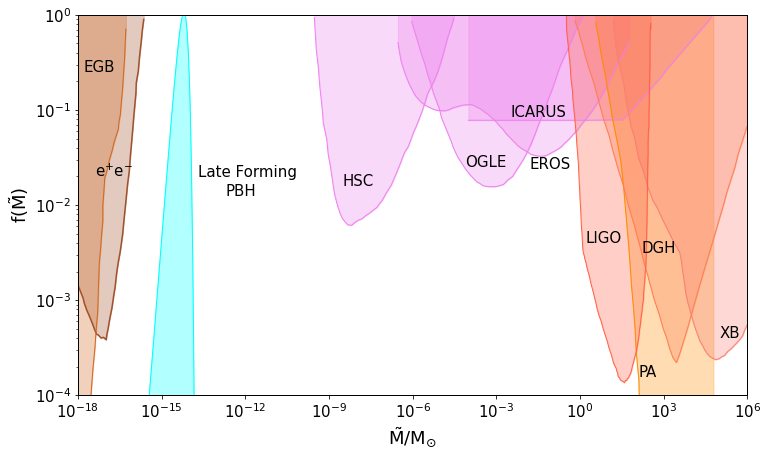}
\caption{Plot of the late forming PBH mass function, $f(\tilde{M})$, shown in cyan, suggests that the PBH density can attribute to the entire dark matter density in the universe. PBH mass function shown in this figure is for $f_{\rm max}(\tilde{M})=1$ (ie., $f_{\rm PBH} \approx 1$) for redshift of formation $z_{\scriptscriptstyle{T}} \sim 10^6$. }
\label{fig:fm_plt_exp}
\end{figure}

We compute the PBH mass function $f(M)$ using Equations \ref{eq:fM} and \ref{eq:Pr_Mh} in which for $\sigma(M_h)$ we use Equations \ref{eq:std}, \ref{eq:xto_m_h} and \ref{eq:pow_spec_exp}. For our calculation we have used $K=11.9$, $\gamma=0.37$ as suggested by \cite{PhysRevD.60.063509} and $\delta_c =2.07$ from \cite{PhysRevD.100.083518}. The corresponding plots is shown in the Figure~\ref{fig:fm_plt_exp}. In this figure all the constraints plot are taken from \cite{kushenko} . For this calculation, we have not taken the whole horizon to be under the influence of the exponential growth but rather taken a small fraction of it, which is basically the ratio between the range of the scalar field, i.e., Compton-scale and the Hubble radius ($(m_\phi/H)^{-1}$).
This is taken into account in the calculation as a factor of $(m_\phi/H)^{-3} \sim 10^{-25}$ to the black hole mass evaluated in our calculation and the plot in Figure~\ref{fig:fm_plt_exp}.
Also, since the Compton scale is smaller than the horizon scale, at any epoch, there will be multiple numbers of Compton patches that will contribute to the formation of PBH. We have incorporated this into the calculation by multiplying a factor of $(m_\phi/H)^{3} \sim 10^{25}$ in the expression of $f(M)$ in equation \ref{eq:fM}. Therefore the modified expression becomes,
\beq\label{eq:fM1}
f(\tilde{M})= \left(\frac{m_\phi}{H}\right)^3\frac{2}{\Omega_{\rm CDM}}\int_{-\infty}^{\infty} Pr(M_h) \frac{\tilde{M}}{\gamma M_h}\tilde{\mu}^{1/\gamma}\left(\frac{M_{\rm eq}}{M_h}\right)^{1/2}d\ln{M_{h}}
\eeq
where, $\tilde{M}=M \left(\frac{m_\phi}{H}\right)^{-3}$  which is defined as the actual black hole mass when a Compton range of collapse is considered and $\tilde{\mu}=\tilde{M}/(K M_h)$.
One can think this treatment is analogous to the modified definition of $\beta$ for non-monochromatic mass function, given in equation \eqref{eq:betaM} as opposed to, in Press-Schechter theory, the factor ${M}/{M_h}$ in $\beta$  never appears, due to assumption of the monochromatic mass function.
Also, hypothetically one can think of the constant Compton scale as our effective  horizon. There is a sharp exponential growth in the power spectrum inside this effective horizon, and it reflects an approximate sharp peak in $f(\tilde{M})$, which we have got in our plot.
    
The plot in figure~\ref{fig:fm_plt_exp} shows a peak $f(\tilde{M})$ for a certain PBH mass for which the PBH contributes to the total dark matter density in the universe. 
Since the Compton scale remains constant with time, the ratio between the Compton scale and the Hubble radius ($(m_\phi/H)^{-1}$) decreases with time. This ratio takes the highest value in our calculation at the beginning of the growth regime. For this reason, the $f(\tilde{M})$ gets the highest contribution from the PBH mass corresponding to this scale. We find that, given enough span for the exponential growth regime ($z_{\scriptscriptstyle T} \sim 10^6$) our model can give $f_{\rm max}(\tilde{M}) = 1$ which is a good approximation to $f_{\rm PBH} = 1$, suggesting that these PBHs can make up for all the dark matter in this case.
Increasing the termination redshift of the growth regime, $z_{\scriptscriptstyle T}$ beyond $\sim 10^6$, decreases the abundance of the PBHs, and hence the corresponding $f_{\rm PBH}$ goes below 1.

\section{Conclusions and Discussion}
\label{sec:conclusion}

In this paper, we present a novel mechanisms for late forming ($z_{\scriptscriptstyle{T}} \sim 10^6$) PBHs. Keeping the scale-invariant inflationary primordial power as it is, here we show that a late phase transition in a fermion-scalar fluid can produce dark matter PBHs due to the formation of early dense halo in radiation dominated era, which evolves into PBH by cooling though scalar radiation.  We also do a detailed analytical calculation of the PBH mass function for these cases and show that, in principle, these PBHs can even make up for almost all of the dark matter. The PBH masses in these models are in the range of $\sim 10^{-16} - 10^{-14} {M}_\odot$.

In standard $\Lambda$CDM cosmology, the growth function being logarithmic in the radiation-dominated era, gravity alone can not clump the matter to collapse into PBH. In this work, we have considered a scenario where the growth rate of the primordial inhomogeneities is exponential. This can happen in a scenario where a force stronger than gravity emerges due to interaction between an additional scalar field and a beyond standard model fermionic particle. The exponential growth of perturbations occurs when the fermionic particles become non-relativistic at around $z\sim 10^6$. This can lead to the formation of numerous low mass dark matter halos, which can further collapse due to scalar radiation and form primordial black holes of similar masses. We also do a detailed calculation of the PBH mass function for this model. 
Though \cite{kushenko} also worked out the PBH fraction for a similar PBH formation mechanism (though in a different model context), there an approximate estimation using Press-Schechter theory was used. In this paper, we have presented a detailed and thorough calculation of f(M) for the first time in our knowledge of early DM halo formation.

There is an allowed mass range for PBH around $10^{-16} - 10^{-10} {M}_\odot$ (sublunar mass range), where the constraints on $f_{\rm PBH}$ are almost non-existent. Though this mass range was previously constrained due to femtolensing, optical microlensing, white-dwarf survival and neutron star capture is no longer constrained in the light of recent studies \cite{2018JCAP...12..005K,2019JCAP...08..031M}. Our model allows the formation of PBH in this window for a $\rm{keV}$ mass fermion and with other viable model parameters- therefore, these PBHs can make up for all the dark matter. In principle we should use non-monochromatic constraints  \citep{2017PhRvD..96b3514C} to compare with our $f(M)$, however since our $f(M)$ turns out to be very narrow (if we approximate our $f(M)$ with a log-normal distribution the corresponding $\sigma$ would be $\lesssim 0.2$) in which case monochromatic constraints will be a good approximation to the non-monochromatic constraints derived using the method provided by \cite{2017PhRvD..96b3514C} (for example the non-monochromatic constraints given for the log-normal distribution $\sigma = 2$ case in Figure~20 of \cite{2021RPPh...84k6902C}).

These PBHs are not expected to either evaporate or even accrete mass significantly in a Hubble time scale \citep{2017JHEAp..13...22R,2018JApA...39....9P} which makes these PBHs a very good candidate for dark matter. Though some works considering the capture of PBHs by white dwarfs and neutron stars suggests strong bounds on the sublunar mass PBHs \citep{2014JCAP...06..026P}, recent studies \citep{2014PhRvD..90j3522D} dispute these bounds on the grounds of recent understanding of dark matter density in globular clusters which is now known to be much lower than assumed in these analyses \citep{2013MNRAS.428.3648I}. Though the constraints on PBH is a rapidly evolving field of research and thus the current PBH abundance constraint over various mass windows should be taken as an order of magnitude estimate, future more detailed studies will be able to put more robust bounds around this mass window.

Also, in our model, the PBHs form through the scalar radiation of dense halos, which are just clumps of matter and have a distribution of initial angular momentum. As a result, these PBH would have considerable spin, which in principle can be detected through future gravity wave observation \cite{Flores:2021tmc}.

\section*{Acknowledgements}
 We are grateful to  Alexander Kusenko for his encouraging and valuable comments on the manuscript. We also thank Yacine Ali Haimoud for reading the manuscript and giving his valuable suggestions. SD and KP  acknowledge SERB, India grant CRG/2019/006147 for supporting this project. We would like to thank ICTS LTPDM workshop on PBH  where a few talks and discussions helped in shaping the work at the initial stage.


\bibliographystyle{aasjournal}
\bibliography{main} 



\appendix
\section{Details of derivations}

\subsection{Relation between smoothing radius $R$ and halo mass $M_h$: Eq~\ref{eq:xto_m_h}}\label{a_1}

Using the Friedmann's equation $H^2={8\pi G \rho}/{3c^2}$, we can write the energy density $\rho$ as,
\beq\label{eq:rho_T}
\rho= \frac{\pi^2 g_{*,\rho}(T)}{30 \hslash^3 c^3} (k_B T)^4
\eeq
where, $k={aH}/{c}$ and the scale-factor $a$ in terms of temperature $T$ of the thermal bath as,
\beq\label{eq:basic}
a=\left(\frac{g_{*,s}(T_0)}{g_{*,s}(T)}\right)^{1/3} \frac{T_0}{T}
\eeq
where $T_0$ is the temperature at $a=1$. Now, the horizon mass can be obtained by,
\beq\label{eq:hor_m_h}
M_h=\frac{4}{3}\pi \left(\frac{c}{H}\right)^3 \frac{\rho}{c^2}= \frac{4\pi c}{3} \frac{\rho}{H^3}
\eeq
from equation \eqref{eq:hor_m_h} and using the Friedmann's equation we get,
\beq\label{eq:hubb}
H=\frac{c^3}{2 G M_h}
\eeq
Now using equations \eqref{eq:rho_T} and \eqref{eq:hubb} we get,
\beq\label{eq:m_t}
M_h=\left(\frac{45 \hslash^3 c^{11}}{16 \pi^3 G^3 k_{B}^4}\right)^{1/2} \left(\frac{1}{g_{*,\rho}(T_h)}\right)^{1/2} \frac{1}{T_{h}^2}
\eeq
Here $T_h$ is the temperature corresponding to the horizon mass $M_h$.
Since we know the length $x$ corresponding to wave number $k$ can be written as, ${1}/{R}={aH}/(2\pi c)$. 
Now using equations \eqref{eq:basic}, \eqref{eq:hubb} and \eqref{eq:m_t}, we get,
\beq\label{eq:x}
\frac{1}{R}=\left(\frac{k_B T_0}{2}\right) \left(\frac{g_{*,s}(T_0)}{g_{*,s}(T_h)}\right)^{1/3} \left(\frac{g_{*,\rho}(T_h)}{45 \pi G \hslash^3 c^3}\right)^{1/4} \frac{1}{\sqrt{M_h}}
\eeq

\subsection{Relation between wave-number $k$ and redshift $z$: 
Eq.~\ref{eq:k(z)_and_z(k)_2}} \label{a_2}

In radiation dominated era, $H={1}/(2t)$ and $a= (4 \Omega_{r,\scriptscriptstyle{0}} H_0^2)^{1/4} \sqrt{t}$. Now using $k=\frac{2\pi}{R}$ and equations \eqref{eq:hubb} and \eqref{eq:x} we finally get,
\beq\label{eq:k(z)_and_z(k)_3}
k=(k_B T_0) \left(\frac{g_{*,s}(T_0)}{g_{*,s}(T_h)}\right)^{1/3} (g_{*,\rho}(T_h))^{1/4} \left(\frac{4\pi^3 H_0^2 \,G\, \Omega_{r,\scriptscriptstyle{0}}}{45\,\hbar^3 c^9}\right)^{1/4} (1+z)
\eeq
or, the $1+z(k)$ can be written as,
\beq\label{eq:k(z)_and_z(k)_4}
1+z(k) = \frac{k}{k_B T_0 }\left(\frac{g_{*,s}(T_0)}{g_{*,s}(T_h)}\right)^{-1/3} g_{*,\rho}^{-1/4}(T_h) \left(\frac{4\pi^3 H_0^2 \,G\, \Omega_{r,0}}{45\,\hbar^3 c^9}\right)^{-1/4}
\eeq


\end{document}